\documentstyle[epsfig]{aipproc}

\begin{document}
\title{Hydrodynamics of Neutron Star Mergers}

\author{Joshua A.~Faber and Frederic A.~Rasio}
\address{Department of Physics, M.I.T., 77 Massachusetts Ave.,
Cambridge, MA 02139}

\maketitle

\begin{abstract}
The final burst of gravitational radiation emitted by coalescing
binary neutron stars carries direct
information about the neutron star fluid, and, in particular,
about the equation of state of nuclear matter at extreme densities. The 
final merger
may also be accompanied by a detectable electromagnetic signal, such as a
gamma-ray burst.
In this paper, we summarize the results of theoretical work done over the 
past decade that has led to a detailed understanding of this hydrodynamic 
merger process for two neutron stars, and we discuss the prospects 
for the
detection and physical interpretation of the gravity wave signals
by ground-based interferometers such as LIGO.
We also present results from our latest post-Newtonian SPH calculations
of binary neutron star coalescence, using up to $10^6$ SPH particles to
compute with higher spatial resolution than ever before the merger of
an initially irrotational system.
We discuss the detectability of our calculated gravity wave signals 
based on power spectra.
\end{abstract}

\section*{Introduction}

Coalescing binary neutron stars (NS) are among the most promising
sources of gravitational radiation that should be detectable by
future generations of gravity wave detectors.  LIGO,
VIRGO, GEO, and TAMA may ultimately not only serve to test the
predictions of the theory of general relativity (GR), but could also yield
important information on the interior structure of neutron
stars, which cannot be obtained directly in any other way.

Compact binary orbits decay through energy losses to gravitational
radiation.
So long as the separation between the two NS is
large, the binary inspiral is well described by a point-mass
treatment, modified to take into account finite-size and
relativistic effects, which act only as small corrections.  At the end
of the inspiral, however, the process is inherently hydrodynamic in
nature.  Large tidal interactions can drive the system into dynamical
instability, at which point a quasi-equilibrium treatment of the binary
breaks down, and a numerical treatment is required to accurately model
the system.  

Essentially all recent calculations agree on the basic picture that
emerges for the final coalescence (see \cite{FR2} and \cite{RSCQG}
for a complete list
of references). As the dynamical stability limit is approached,
typically at separations of $r=3-4 R_{NS}$, depending on the choice of
parameters, the NS undergo a rapid radial plunge and merge in
no more than a few rotation periods, much more quickly than would be
predicted by a point-mass formula.  In many cases, especially for
binaries assumed to be initially synchronized, mass shedding
sets in immediately after the stars first make contact.  Material with
a high specific angular momentum located in the outer regions of each
NS is shed through the outer Lagrange points of the system, forming
spiral arms that encircle the merger remnant left in the
center as the NS cores merge.
Eventually, the spiral arms also
merge into a nearly axisymmetric torus around the dense inner
core.  For a stiff equation of
state (EOS), a core with a significant
ellipsoidal (triaxial) deformation can be maintained, and the configuration
keeps radiating gravity waves well after the merger is completed.  Softer 
EOS cannot
support such a configuration stably, and any remnant produced will
relax on a dynamical timescale toward a spheroidal (axisymmetric)
configuration, which produces a negligible amount of gravity waves.  

The previous statements assume that the merger remnant formed is
stable against gravitational collapse to a black hole.  
Unfortunately, Newtonian calculations are incapable of demonstrating
such an effect.  Post-Newtonian (PN) simulations can produce configurations
that are unstable against collapse, but they are inherently unreliable
because in conditions of strong gravity the basic assumptions of the
PN expansion break down.  Early full GR
calculations indicate that merger remnants may very well be stable
against collapse, so long as the EOS is assumed to be
stiff enough \cite{Shi}.  The mass of the remnant should
be nearly twice the mass of a single NS, which is generally taken to
be $M_{NS}\approx 1.4-1.5 M_{\odot}$, and thus well over the maximum
mass for a single, nonrotating NS.  However, the remnants formed in 
binary coalescence
are very rapidly and differentially rotating, which
can increase the maximum stable mass to a much larger value 
\cite{Baum,RasESO}.

\section*{Binary NS Coalescence Calculations}

Nakamura and collaborators \cite{Nak,SON} were the first group to perform 3-D
hydrodynamic calculations of binary NS coalescence, using an Eulerian
grid-based code.  Rasio and Shapiro \cite{RS} used the Lagrangian SPH
(Smoothed Particle Hydrodynamics) method to calculate gravitational
wave forms from binary coalescence events.  Calculations performed
since, using both Eulerian methods \cite{NT,SWC,RJS,RJTS,RRJ}  
and SPH \cite{Zhu,Dav,Ros1,Ros2} have
focused on several aspects of the problem, including the effects of 
different initial spins, mass ratios, NS EOS, and NS masses.
Some groups have incorporated
treatments of the nuclear physics involved in the merger
\cite{RJS,RJTS,RRJ,Dav,Ros1,Ros2} 
in order to study coalescing NS binaries as possible gamma-ray
burst sources, and as possible birthplaces for r-process
elements.

Much of the early work on coalescing NS binaries
assumed Newtonian gravity for simplicity. Later studies added a
treatment of the radiation reaction, which is responsible for driving
the system towards coalescence, either by adding a frictional drag
term to model point-mass inspiral \cite{Zhu,Dav,Ros1,Ros2}, or by an
exact PN treatment \cite{RJS,RJTS,RRJ}.  
In essence, 2.5PN radiation reaction terms (which scale like $1/c^5$)
are added onto a Newtonian framework, but all lower-order
non-dissipative terms are ignored.  Unlike adding a frictional
drag term which dissipates energy according to the
point-mass prediction, the lowest-order treatment of the radiation
reaction allows for its effects to be included throughout the entire
calculation, including the period after the merger remnant has formed.
Unfortunately, however,
Newtonian gravity is known to be a poor description of the physical
problem at hand. Even NS with stiff EOS generate strong gravitational fields.
During the
final moments before merger, the velocities found in the system also
become relativistic.  Thus, the hydrodynamics of the actual
coalescence can only be calculated properly by taking into account
GR effects.

The Newtonian limit also fails to describe accurately the
onset of dynamical instability. PN effets combine nonlinearly with
finite-size fluid effects
and this can dramatically increase the critical binary separation (and
thus lower the frequency) at which dynamical instability sets in. 
Indeed, the quasi-equilibrium description applies so long as 
\begin{equation}
\left(\frac{dE}{dr}\right)_{equil}>\left(\frac{dE}{dt}\right)_{GW}
\left(\frac{dr}{dt}\right)_{infall}^{-1},
\end{equation}
where $E_{equil}(r)$ is the energy of an equilibrium binary
configuration at a given separation $r$.  
Equilibrium sequences have been calculated for both synchronized 
and irrotational 
binaries in Newtonian gravity \cite{LRS}, 
PN gravity \cite{LomRS}, and recently in full GR
\cite{GRSync,GRIrr} (see also Baumgarte, in this volume). 
It is generally found
that the energy of NS binary configurations reaches a minimum at
some critical separation, defining the innermost stable circular orbit (ISCO).
Relativistic terms move the ISCO to larger separations, and also
reduce the slope of the energy curve
just outside the ISCO.  Thus, the assumption of
quasi-equilibrium, which is used to set up the initial configuration
of the binary system, breaks down at a much larger separation than a
Newtonian calculation would predict. In addition, NS binaries 
will already have developed significant infall velocities as they pass through
the ISCO calculated for systems in strict equilibrium.

Several groups have been working on full GR
calculations of binary NS mergers, but only preliminary results have
been reported so far \cite{GR}.  Proving to be particularly difficult is
the extraction of wave forms from the boundaries of large 3-D grids, since
extending the grids into the true wave zone would be too expensive
computationally.
The middle ground between the Newtonian treatments and full GR lies in
PN hydrodynamic calculations of binary mergers.  The
authors \cite{FR2,FR1}, as
well as Ayal et al. \cite{Ayal}, have constructed a PN
SPH code, described below, for calculating binary mergers.  While it
too serves only as an approximation to the proper physics which must
go into a realistic calculation, the results do provide insight into
the relativistic effects that simple Newtonian intuition fails to
handle correctly.  Additionally, they should serve as valuable checks
for future full GR calculations, lending
confidence to wave form predictions, and indicating to some degree the
difference between real relativistic effects and numerical
instabilities.   

\section*{Post-Newtonian SPH}

Our Post-Newtonian calculations use a formalism adapted from that of
Blanchet, Damour, and Schaefer \cite{BDS}.  It includes all
first-order (1PN) terms of GR, as well as the
lowest-order radiation reaction terms (2.5PN).  The latter are
important because they provide the energy dissipation mechanism which
drives the binary system toward dynamical instability and
coalescence.  Calculating various PN quantities requires the solution
of seven additional Poisson equations for 1PN terms and an additional
Poisson equation for the radiation reaction, all of which have compact
support and thus can be solved by similar methods as the Newtonian
gravitational potential.
All hydrodynamic quantities in the formalism have been
converted from a grid-based Eulerian approach to a particle-based
Lagrangian approach, where particles represent
distributions
of matter in space defined by a spherically symmetric smoothing
kernel (rather than point-like objects).  Pressure forces and other 
interactions between particles are
handled by summing over neighboring particles.
Poisson equations are solved by translating particle-based source term
quantities onto a grid, and solving via FFT-based 
convolution methods \cite{FR1}.

Unfortunately, the consistent use of physically realistic NS parameters
is impossible in this PN formalism.  The compactness of a NS is given by
\begin{equation}
\frac{GM}{Rc^2}=0.20\left(\frac{M}{1.5M_{\odot}}\right)
\left(\frac{10~{\rm km}}{R}\right),
\end{equation}
and is typically assumed to fall near
$GM/Rc^2\approx0.15$ for a reasonable choice of EOS.  Since several of the
coefficients of the 1PN terms can be quite large, especially for stiff
EOS, we find that in many cases the 1PN corrections are larger than
the Newtonian terms.  Thus, we adapt the formalism, reducing the
magnitude of the 1PN corrections by a factor of three (in effect
decreasing the mass of the NS to $M_{1PN}=0.5M_{\odot}$), but treating
all the radiation reaction terms, of significantly smaller magnitude, 
at full strength.  In essence, we employ a different speed of light
for 1PN and 2.5PN terms.
By comparison with result including
radiation reaction but no 1PN terms whatsoever, we believe we
can extrapolate in a qualitative sense toward the proper physically
realistic case.  
	By comparison, self-consistent PN calculations performed for NS with
artificially small masses \cite{SON,Ayal}
have the advantage that they can be directly
compared to full GR calculations, for which there can be no
separation of relativistic terms into separate orders.  Unfortunately,
these calculations also suffer from a drastic and completely artificial
reduction of the
radiation reaction effects, which scale
like $M^{2.5}$.  This produces a significant delay in the onset of
dynamical instability past the point where it would be encountered
for a physical set of parameters, and can lead to a qualitatively 
incorrect description of the subsequent merger.

\section*{Irrotational Binary Coalescence}

Our most detailed calculation performed to date uses $N=500,000$
particles per NS, corresponding to the highest spatial resolution 
ever for a binary coalescence calculation.  The spatial resolution 
(smoothing length) achieved in the central
regions of the stars is $h\approx 0.03 R_{NS}$.  
The calculation was performed using an irrotational initial condition.
This is generally thought to be the most realistic case since the 
viscous tidal locking timescale for two NS is expected to be
considerably longer than the inspiral timescale \cite{BC}.
Corotating (tidally locked) systems are
motionless in a frame corotating with the binary, which allows
for the use of relaxation techniques that can give a very accurate
equilibrium initial condition (thus nearly completely eliminating
spurious oscillations
around equilibrium during the early phases of the dynamical
evolution). Instead, for our irrotational calculation, we model
the initial density and velocity profile of the NS as tidally stretched
ellipsoids, with parameters drawn from the PN equlibrium calculations
of Lombardi, Rasio, and Shapiro \cite{LomRS}.

Since all NS in relativistic binary systems are expected to have
 masses that lie
within a very narrow range \cite{TC}, 
our calculation uses equal-mass NS.  As
the NS EOS is still poorly constrained, we choose a
simple $\Gamma=3$ polytropic EOS, i.e., the pressure is given in terms
of the rest-mass density by $P=k\rho_*^3$.  Stiff EOS, such as this 
one, are
capable of maintaining a long-lived ellipsoidal deformation after the
binary merger, with a gravity wave signal that persists on a timescale
much longer than the merger timescale \cite{RS2}.

\begin{figure}[t!]
\centerline{\epsfig{file=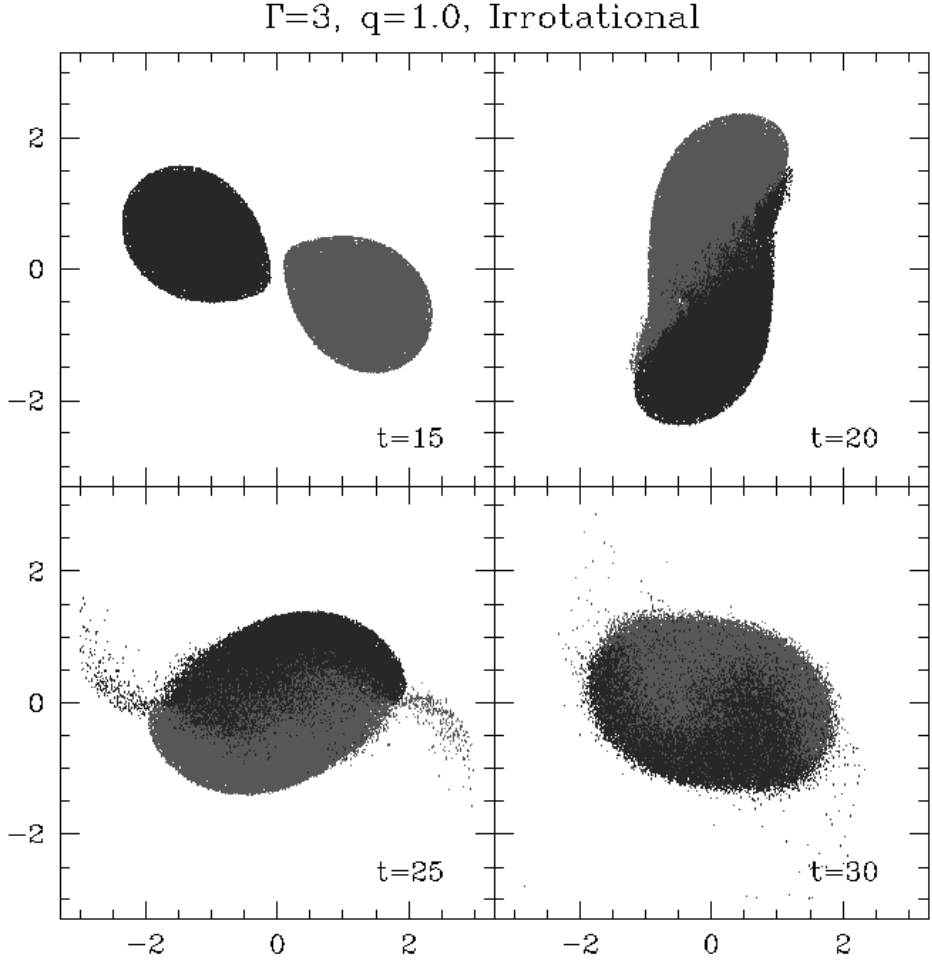,height=5.5in,width=5.5in}}
\vspace{10pt}
\caption{Final merger of two identical $\Gamma=3$ polytropes with
an irrotational initial condition.
SPH particles are projected onto the
equatorial plane of the binary. The orbital rotation is counterclockwise.
Spatial coordinates are given in  units 
of the NS radius $R$.  Times are given in
units of the dynamical timescale of the system, which here is $t_D=0.07{\rm
ms}=1$.  The orbital rotation is in the counterclockwise direction.}
\label{fig1}
\end{figure}

Particle plots showing the evolution of the equal-mass irrotational
binary system are shown in Fig.~\ref{fig1}.  
We see that immediately prior to merger a large tidal lag angle
develops.  The inner edge of each NS leads the axis connecting the
centers of mass of the binary components, and the outer regions lag
behind.  This effect is seen even in Newtonian simulations, but is
greatly enhanced by the addition of 1PN correction terms.   When first
contact is made, a long vortex sheet forms at the interface between the 
two stars.
Unlike the case of synchronized binaries, we do not see significant
mass shedding from the outer Lagrange points of the system.  The
rotational speed of particles on the outer half of each NS is reduced
in the irrotational case with respect to the synchronized case, and
such particles remain bound and form the outer regions of the eventual
merger remnant.  At $t=25$ we do see some hint of mass shedding, but
not via the mechanism described above.  Particles which have travelled
the length of the vortex sheet and retained significant velocities end
up being shed from the leading edge of the vortex sheet.  It is
important to note, though, that the amount of mass shed is extremely
small, much less than 1\% of the total mass, and that
the velocities of the particles ejected are not sufficient to escape
the gravitational potential of the remnant.  We thus expect
them to form an extremely tenuous halo around the central core.  At
late times, we see the formation of a remnant containing
essentially all the mass that was orignally present in the system.  As
we are using a stiff EOS, the ellipsoidal deformation of the remnant
is relatively large, and persists for late times after the merger.

\begin{figure}[t!]
\centerline{\epsfig{file=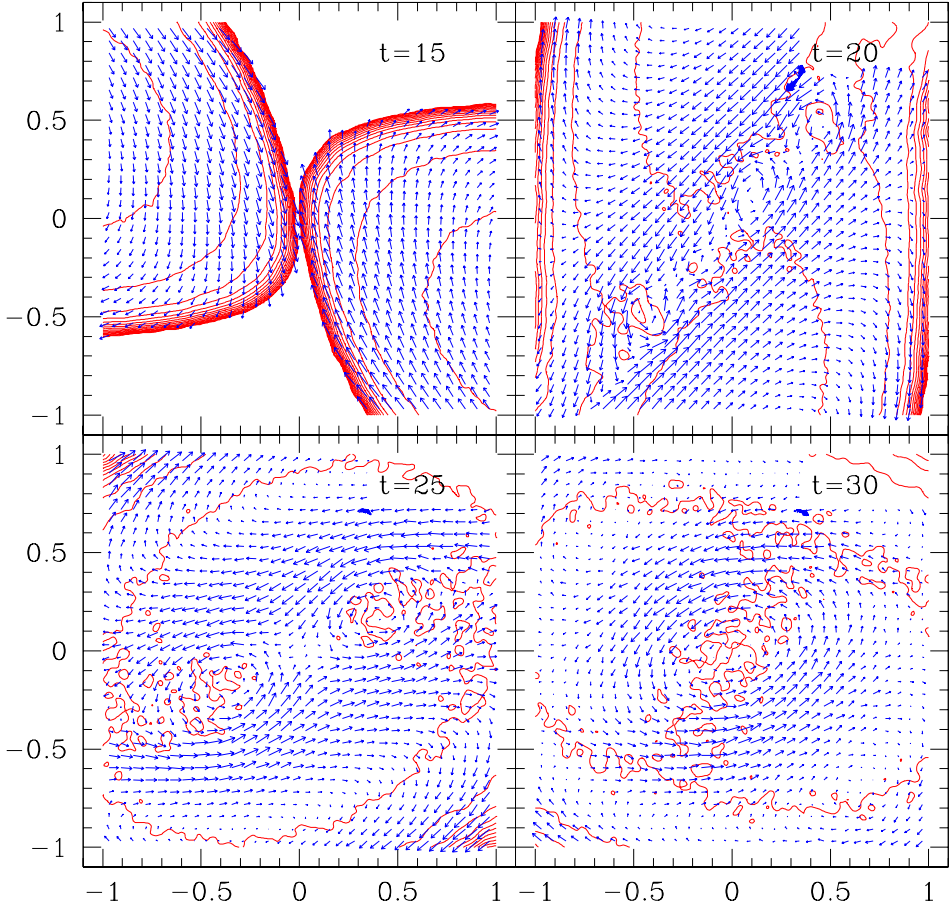,height=5.5in,width=5.5in}}
\vspace{10pt}
\caption{Density contours and velocities along the equatorial plane
in the corotating frame of the binary, for the same times as
in Fig.~\protect\ref{fig1}.}
\label{fig2}
\end{figure}

Density contours and velocity profiles in the equatorial plane of the
binary are shown in Fig.~\ref{fig2}.  Velocities are shown in a frame 
corotating with the material.
We see that the initially counterstreaming
surfaces of the NS produce a vortex sheet, which is Kelvin-Helmholtz
unstable. 
Large vortices develop along the surface of contact, both in the center of
the forming merger remnant and also at a separation which seems to be
roughly consistent with the misalignment of the leading edges of each
NS, or about $r=0.5R$.  As the merger proceeds, we see that the
vortices remain coherent from $t=20-25$, mixing material which was
orignally located along the inner parts of each NS.  All the while,
the cores of the respective NS continue to inspiral toward the center
of the merger remnant, until by $t=30$ they have formed a single
core, the vortices having merged together.  This
produces a characteristic differentially rotating pattern, with the
center of the remnant spinning approximately twice as fast as the
outer regions.  

\section*{Gravity Wave Signals and Spectra}

We calculate the gravity wave signal for our mergers in the quadrupole
approximation.  The gravity wave strain $h$ seen by an observer located a
distance $d$ from the center of mass of the system along the rotation
axis is given for the two polarizations by
\begin{eqnarray}
c^4(dh_+)&=&\ddot{Q}_{xx}-\ddot{Q}_{yy}\\
c^4(dh_{\times})&=&2\ddot{Q}_{xy}
\end{eqnarray}
where $\ddot{Q}$, the second time derivative of the
quadrupole moment tensor, is given in SPH terms by 
\begin{equation}
\ddot{Q}_{ij}=\sum_b m_b(v_i^{(b)} v_j^{(b)}+x_i^{(b)}\dot{v}_j^{(b)}+
x_j^{(b)}\dot{v}_i^{(b)})
\end{equation}
where the summation is taken over all particles in the calculation.

\begin{figure}[t!]
\centerline{\epsfig{file=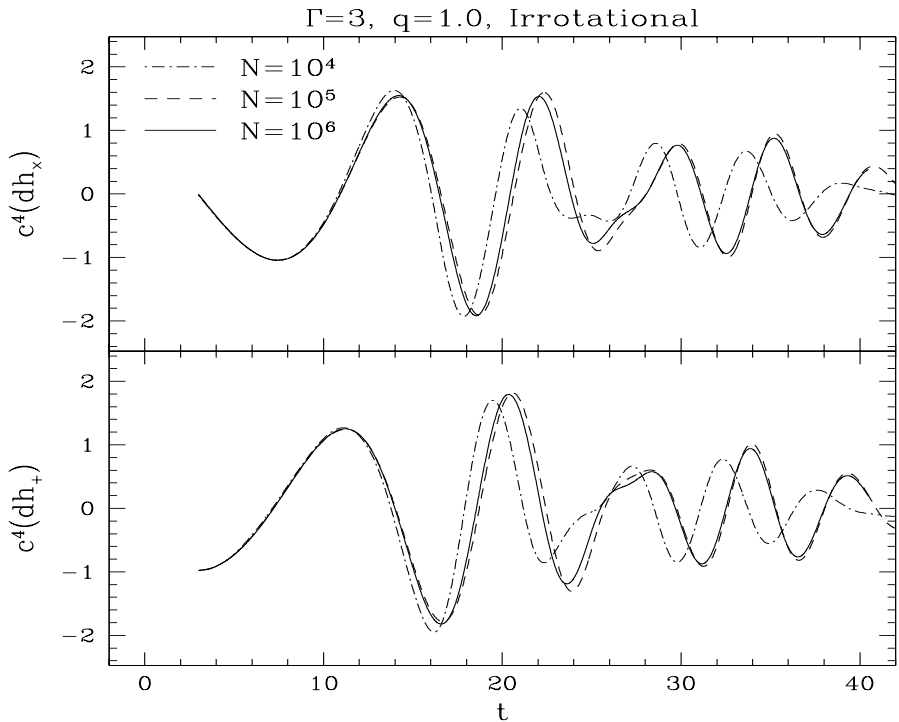,height=4.0in,width=5.0in}}
\vspace{10pt}
\caption{Gravity wave signals calculated for coalescences with the
same initial parameters but different numerical resolutions.  The
solid, dashed, and dot-dashed lines correspond to runs with $10^6$,
$10^5$, and $10^4$ SPH particles, respectively.}
\label{fig3}
\end{figure}

In Fig.~\ref{fig3}, we show the gravity wave signals in both
polarizations for the irrotational run described above, as well as for
runs with $N=50,000$ particles and $N=5,000$ particles per NS.  
It is immediately apparent that the lowest resolution run shows
significant discrepancies from the other two, which agree with each
other quite well over the entire time history of the merger.  This is
a welcome result, given that the vortex sheet appearing
at the contact surface is Kelvin-Helmholtz unstable on all
wavelengths, including those much smaller than our numerical
resolution.  
Calculations performed at different resolutions do show
subtle differences in the exact location and size of the vortices.  
It is important to note, however, that it is
the outer regions of the star, at lower density, that supply material
to the vortex sheet.  The high density cores of the two NS
inspiral during the entire process, and provide the dominant
component of the quadrupole moment and thus the gravity wave signal.
The path traced out by the NS cores depends sensitively on
gravitational forces and properties of the fluid, such as the
EOS, but proves to be remarkably insensitive to the
details of the flow in the ``turbulent'' boundary region.  
The conclusion to be drawn is
that numerical convergence for a given set of initial conditions and
physical assumptions is possible without requiring excessive
computational resources, even for this difficult problem involving
small-scale instabilities.

Because the gravity wave signals expected to be seen from binary NS
mergers are extremely close to the sensitivity limits of 
ground-based interferometers, it is important to
identify which features in the power spectrum of the gravitational
radiation will yield the most information on the physical
parameters of NS.  Following the method of Zhuge et al. \cite{Zhu},
we compute the gravity wave power spectrum per unit frequency interval as
\begin{equation}
\frac{dE_{GW}}{df}=\frac{c^3}{G}\frac{\pi}{2}(4\pi r^2)f^2\left\langle
|\tilde{h}_{+}(f)|^2+|\tilde{h}_{\times}(f)|^2\right\rangle.
\end{equation}
Before calculating the power spectra from our simulations, we
add a component representing a point-mass inspiral matched to the
beginning of our hydrodynamic merger wave form.  
This produces a spectrum with $dE/df\propto f^{1/3}$ for point-mass
inspiral at low frequencies.  
In essence, what we measure here is the number of orbits
spent around a given frequency, weighted by the amplitude of the
emission.  We identify three frequencies of special interest.  The
frequency at which dynamical instability sets in is labeled $f_{dyn}$.
The frequency at the peak of the gravity wave
luminosity is labeled $f_{peak}$.  Last, the characteristic frequency
of gravity wave emission for the merger remnant at late times in the
calculation is labeled $f_{osc}$.  

\begin{figure}[t!]
\centerline{\epsfig{file=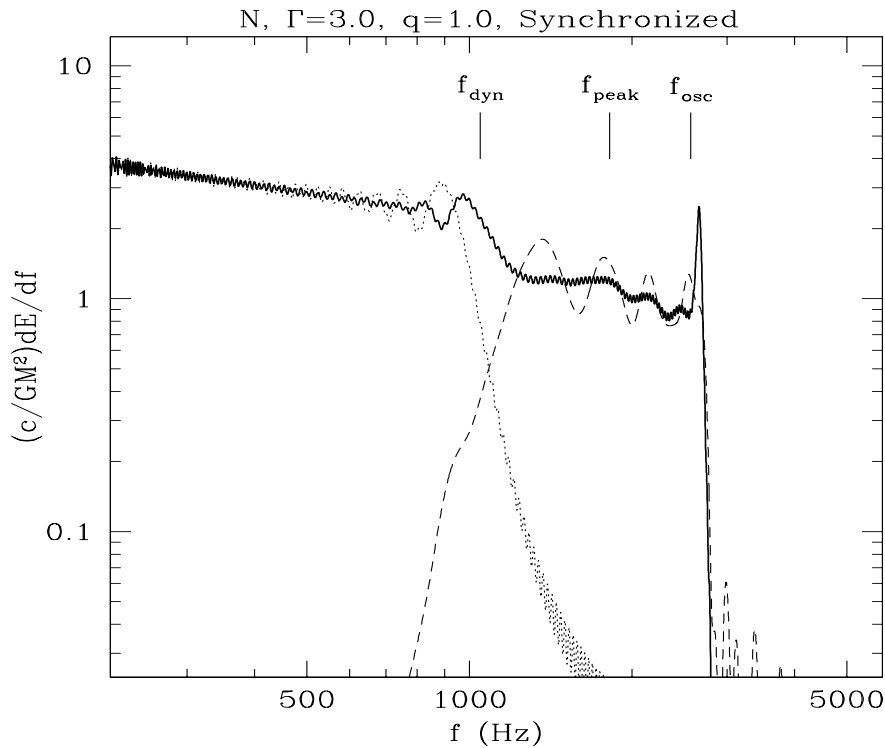,height=4.0in,width=5.0in}}
\vspace{10pt}
\caption{Power spectrum calculated from a Newtonian coalescence calculation.  
The
dotted and dashed lines represent the contributions of the point-mass
inspiral and our calculated gravity wave signal for the hydrodynamic merger, 
respectively.  The solid line represents the total power.}
\label{fig4}
\end{figure}

In Fig.~\ref{fig4}, we show the
gravity wave power spectrum for a Newtonian calculation with a
synchronized initial condition, which
includes radiation reaction effects but contains no 1PN terms.  
At low frequencies, we see the power-law behavior from the point-mass
inspiral, with no contribution whatsoever from our calculated signal.
At the dynamical instability frequency, we see a slight decrease in
the gravity wave power, since the rapid plunge causes the binary to
sweep up faster through a range of characteristic frequencies.  The
gravity wave power shows a plateau near the peak emission frequency,
when the effects of the rapid infall are balanced by the large
increase in gravity wave amplitude.  At higher frequencies, there is 
another slight dip in emission, followed by a sharp peak marking 
the oscillation
frequecncy of the merger remnant.  With more careful handling of the
late-time behavior of the system, we expect the peak to remain prominent,
but our calculation most likely overemphasizes the coherence of the
signal.  

\begin{figure}[t!]
\centerline{\epsfig{file=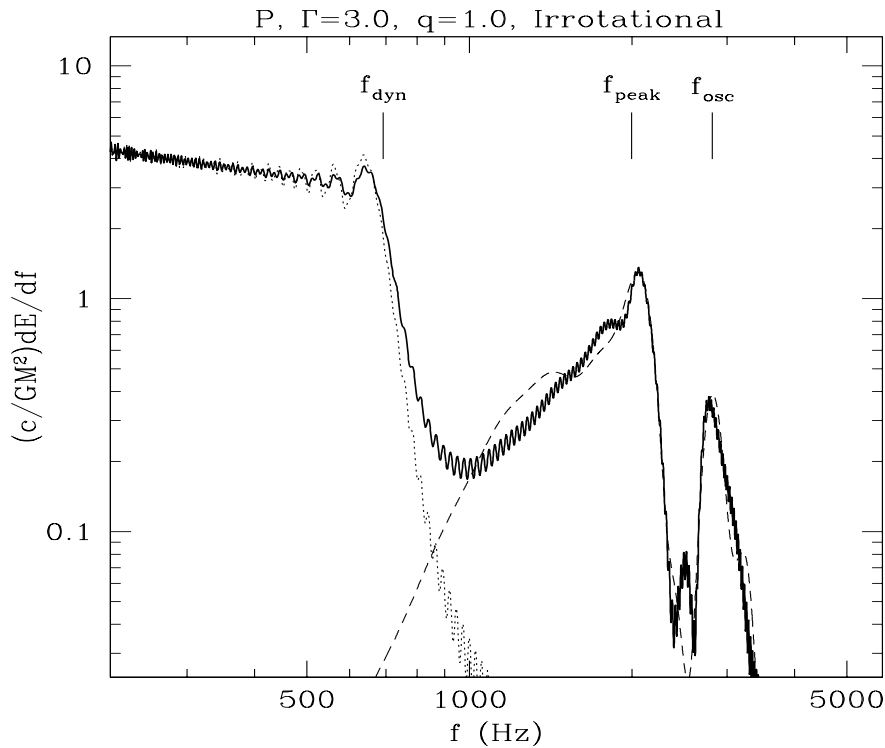,height=4.0in,width=5.0in}}
\vspace{10pt}
\caption{Power spectrum from a Post-Newtonian coalescence calculation.
Conventions are as in the previous figure.}
\label{fig5}
\end{figure}

A strikingly different  power spectrum is obtained from
our Post-Newtonian, irrotational merger, as shown in Fig.~\ref{fig5}.
The most significant difference between the two concerns the limit of
dynamical instability.  Newtonian gravity is strengthened by the
addition of relativistic effects, which moves the dynamical stability
limit to larger separations (and thus smaller frequencies).
Additionally, the final plunge of the two NS is much faster, which
significantly reduces the power in the region between $f_{dyn}$ and
$f_{peak}$.  At $f_{peak}$, the power is smaller than in the Newtonian
case, but the signal is much more sharply defined.  Similarly, the
oscillation of the remnant at late times leaves a well defined imprint
on the power spectrum, but the amplitude of the peak is approximately
an order of magnitude lower than what is found in a Newtonian
calculation.  

The frequencies of the two peaks seen in the spectrum, representing
peak emission and the remnant oscillations, do give a strong clue to
the nature of the NS EOS.  While the frequency of peak oscillation is
essentially the same in all our simulations, the width of the peak is
seen to be strongly dependent on the EOS.  The softer $\Gamma=2$ EOS
shows a broad peak of emission in the frequency range $f\sim
1500-3000~{\rm Hz}$, whereas the stiffer $\Gamma=3$ EOS calculations
have a peak much more focused around $f=1800-2200~{\rm Hz}$,
regardless of the initial spins.  The remnant oscillations break this
degeneracy.  The oscillation frequency for the irrotational run is
almost $15\%$ less than that of the synchronized run with the same
choice of EOS.  The stiffer EOS also results in a more rapid
oscillation than the softer one, although the frequencies are
relatively similar.

In general, several trends can be recognized regarding the strength of
gravity wave emission from NS binary coalescence.  The
large dip at the dynamical instability limit is a general feature of
PN calculations, regardless of the choice of initial spins
or the EOS.  It should be regarded as a consequence
of the stronger gravity present in the PN systems.  Additionally,
PN simulations generally show similar amplitudes to their
Newtonian counterparts near the characteristic frequency of peak
emission, but significantly less power at higher
frequencies, since the effect of strong gravity seems to be a
quenching of gravity wave emission after the initial peak.  Soft
EOS generally show greatly reduced gravity wave emission
at high frequencies, since they cannot support a stable, radiating,
ellipsoidal configuration, and on relatively short timescales will
produce a nearly spheroidal, non-radiating remnant.  Finally, for
binary systems with unequal-mass components, the magnitude of the
gravity wave emission is strongly correlated with the mass ratio $q$
\cite{FR2,RS2}.
Because the primary in such systems generally remains relatively
undisturned, whereas the secondary is tidally disrupted and accreted
onto the primary, a large component of the matter essentially does not
contribute to the gravity wave signal.  Thus, even if NS masses do not
typically lie within a narrow range, there should be a strong bias
observationally toward detection of nearly equal-mass systems.

\acknowledgments
This work was supported in part by NSF Grants AST-9618116 and PHY-0070918
and NASA ATP Grant NAG5-8460. 
F.A.R.\ was supported in part by an Alfred P.\ Sloan Research Fellowship.
The computations were 
supported by the National Computational Science Alliance under grant
AST980014N and utilized the NCSA SGI/CRAY Origin2000.


\begin{references}
\bibitem{FR2}Faber, J.A., Rasio, F.A., and Manor, J.B., {\it
Phys. Rev. D}, accepted, gr-qc/9912097.
\bibitem{RSCQG} Rasio, F.A., and Shapiro, S.L., {\it
Class. Quant. Grav.} {\bf 16}, R1-R29 (1999).
\bibitem{Shi}Shibata, M., and Uryu, K., {\it Phys. Rev. D} {\bf 61},
064001 (2000).
\bibitem{Baum}Baumgarte, T.W., Shapiro, S.L., and Shibata, M., {\it
Astrophys. J. Lett.} {\bf 528} L29-L32 (2000).
\bibitem{RasESO} Rasio, F.A., ``The Final Fate of Coalescing Binary
Neutron Stars: Collapse to a Black Hole?'' to appear in {\it Black
Holes in Binaries and Galactic Nuclei}, edited by L. Kaper, E.P.J. van
den Heuvel, and P.A. Woudt, ESO Press.   
\bibitem{Nak}Oohara, K., and Nakamura, T., {\it Prog. Theor. Phys.}
{\bf 82}, 535-554 (1989); {\it ibid.} {\bf 83}, 906-940 (1990);
Nakamura, T. and Oohara, K., {\it ibid.} {\bf 82}, 1066-1083 (1989);
{\it ibid.} {\bf 86}, 73-88 (1991). 
\bibitem{SON} Shibata, M., Oohara, K., and
Nakamura, T., {\it Prog. Theor. Phys.} {\bf 88}, 1079-1095 (1992);
{\it ibid.} {\bf 89}, 809-819 (1993).
\bibitem{RS} Rasio, F.A., and Shapiro, S.L., {\it Astrophys. J.} {\bf
401}, 226-245 (1992); {\it ibid.} {\bf 438}, 887-903 (1995).
\bibitem{RS2} Rasio, F.A., and Shapiro, S.L., {\it Astrophys. J.} {\bf
432}, 242-261 (1994).  
\bibitem{NT} New, K.C.B., and Tohline, J.E., {\it Astrophys. J.} {\bf
490}, 311-327 (1997). 
\bibitem{SWC} Swesty, F.D., Wang, E.Y.M., and Calder, A.C., {\it
Astrophys. J.} {\bf 541}, 937-958 (2000).
\bibitem{RJS} Ruffert, M., Janka, H.-Th., and Sch\"{a}fer, G., {\it
Astron. Astrophys.} {\bf 311}, 532-566 (1996).
\bibitem{RJTS} Ruffert, M., Janka, H.-Th., Takahashi, K., and
Sch\"{a}fer, G., {\it Astron. Astrophys.} {\bf 319}, 122-153 (1997).  
\bibitem{RRJ} Ruffert, M., Rampp, M., and Janka, H.-Th., {\it
Astron. Astrophys.} {\bf 321}, 991-1006 (1997).
\bibitem{Zhu} Zhuge, X., Centrella, J., and McMillan, S., {\it
Phys. Rev. D} {\bf 50}, 6247-6261 (1994); {\it ibid.} {\bf 54},
7261-7277 (1996).
\bibitem{Dav} Davies, M.B. et al., {\it Astrophys. J.} {\bf 431},
742-753 (1994).
\bibitem{Ros1} Rosswog, S. et al., {\it Astron. Astrophys.} {\bf 341},
499-526 (1999).
\bibitem{Ros2} Rosswog, S. et al., {\it Astron. Astrophys.} {\bf 360},
171-184 (2000).
\bibitem{LRS} Lai, D., Rasio, F.A., and Shapiro, S.L., {\it
Astrophys. J. Lett.} {\bf 406}, L63-L66 (1993); {\it
Astrophys. J. Suppl.} {\bf 88}, 205-252 (1993); {\it Astrophys. J.}
{\bf 420}, 811-829 (1994); {\it ibid.} {\bf 423}, 344-370 (1994); {\it
ibid.} {\bf 437}, 742-769 (1994).
\bibitem{LomRS} Lombardi, J.C., Rasio, F.A., and Shapiro, S.L., {\it
Phys. Rev. D} {\bf 56}, 3416-3438 (1997).
\bibitem{GRSync} Baumgarte, T.W. et al., {\it Phys. Rev. Lett.} {\bf
79} 1182-1185 (1997).
\bibitem{GRIrr} Bonazzola, S., Gourgoulhon, E., and Marck,  J.-A.,
{\it Phys. Rev. Lett.} {\bf 82} 892-895 (1999).
\bibitem{GR} Baumgarte, T.W., Hughes, S.A., and Shapiro, S.L., {\it
Phys. Rev. D.} {\bf 60}, 087501 (1999); Shibata, M., {\it
Phys. Rev. D} {\bf 60}, 104052 (1999).
\bibitem{FR1} Faber, J.A. and Rasio, F.A., {\it Phys Rev. D} {\bf 62},
064012 (2000).
\bibitem{Ayal} Ayal, S. et al., {\it Astrophys. J.}, accepted,
astro-ph/9910154. 
\bibitem{BDS} Blanchet, L., Damour, T., and Sch\"{a}fer, G., {\it
Mon. Not. Roy. Astron. Soc.} {\bf 242}, 289-305 (1990).
\bibitem{BC} Bildsten, L., and Cutler, C., {\it Astrophys. J.} {\bf
400}, 175-180 (1992).
\bibitem{TC} Thorsett, S.E., and Chakrabarty, D., {\it Astrophys. J.}
{\bf 512}, 288-299 (1999).
\end{references}
\end{document}